\documentclass[onecolumn,prl, preprint, superscriptaddress]{revtex4-2}
\usepackage{bm}
\usepackage[colorlinks=true,linkcolor=blue,citecolor=blue]{hyperref}
\usepackage{times}
\usepackage{amsmath}
\usepackage{amssymb}
\usepackage{amsthm}
\usepackage{amsfonts}
\usepackage{enumerate}
\usepackage{latexsym}
\usepackage{ifpdf}
\newcommand{\beq}{\begin{equation}}
\newcommand{\eeq}{\end{equation}}
\usepackage{graphicx}
\usepackage{makeidx}
\usepackage{color}

\usepackage[version=4]{mhchem}
\usepackage{textcomp,mathcomp}

\begin{document}

\title{Site-selective polar compensation of Mott electrons in a double perovskite heterointerface}

\author {Nandana Bhattacharya}
\affiliation  {Department of Physics, Indian Institute of Science, Bengaluru  560012, India}
\author {Arpita Sen}
\altaffiliation{Contributed equally}
\altaffiliation  {Currently at Department of Physics, School of Natural Sciences, Shiv Nadar Institution of Eminence, Greater Noida, Gautam Buddha Nagar, Uttar Pradesh 201314, India}
\affiliation  {Theoretical Sciences Unit, Jawaharlal Nehru Centre for Advanced Scientific Research, Jakkur P.O., Bangalore 560 064, India}

\author {Ke Qu}
\altaffiliation{Contributed equally}
\affiliation{Key Laboratory of Polar Materials and Devices, East China Normal University Shanghai, Shanghai 200241, China}
\author {Arijit Sinha}
\altaffiliation{Contributed equally}
\affiliation  {Theoretical Sciences Unit, Jawaharlal Nehru Centre for Advanced Scientific Research, Jakkur P.O., Bangalore 560 064, India}
\author {Ranjan Kumar Patel}
\affiliation  {Department of Physics, Indian Institute of Science, Bengaluru 560012, India}
\author {Siddharth Kumar}
\affiliation{Department of Physics, Indian Institute of Science, Bengaluru 560012, India}
\author {Jianwei Zhang}
\affiliation{Key Laboratory of Polar Materials and Devices, East China Normal University Shanghai, Shanghai 200241, China}
\author {Prithwijit Mandal}
\affiliation  {Department of Physics, Indian Institute of Science, Bengaluru 560012, India}
\author {Suresh Chandra Joshi}
\affiliation  {Department of Physics, Indian Institute of Science, Bengaluru 560012, India}
\author {Shashank Kumar Ojha}
\affiliation{Department of Physics, Indian Institute of Science, Bengaluru 560012, India}
\author {Jyotirmay Maity}
\affiliation{Department of Physics, Indian Institute of Science, Bengaluru 560012, India}
\author{Zhan Zhang}
\affiliation {Advanced Photon Source, Argonne National Laboratory, Lemont, Illinois 60439, USA}
\author{Hua Zhou}
\affiliation {Advanced Photon Source, Argonne National Laboratory, Lemont, Illinois 60439, USA}
\author{Fanny Rodolakis}
\affiliation {Advanced Photon Source, Argonne National Laboratory, Lemont, Illinois 60439, USA}
\author {Padraic Shafer}
\affiliation  {Advanced Light Source, Lawrence Berkeley National Laboratory, Berkeley, California 94720, USA}
\author{Christoph Klewe}
\affiliation  {Advanced Light Source, Lawrence Berkeley National Laboratory, Berkeley, California 94720, USA}
\author{John William Freeland}
\affiliation {Advanced Photon Source, Argonne National Laboratory, Lemont, Illinois 60439, USA}
\author {Zhenzhong Yang}
\email{zzyang@phy.ecnu.edu.cn}
\affiliation{Key Laboratory of Polar Materials and Devices, East China Normal University Shanghai, Shanghai 200241, China}
\author {Umesh Waghmare}
\affiliation  {Theoretical Sciences Unit, Jawaharlal Nehru Centre for Advanced Scientific Research, Jakkur P.O., Bangalore 560 064, India}
\author {Srimanta Middey}
\email{smiddey@iisc.ac.in}
\affiliation  {Department of Physics, Indian Institute of Science, Bengaluru 560012, India}

\begin{abstract}

 Double perovskite oxides (DPOs) with two transition metal ions ($A_2$$BB^\prime$O$_6$) offer a fascinating platform for exploring exotic physics and practical applications. Studying these DPOs as  ultrathin epitaxial thin films on single crystalline substrates can add another dimension to engineering electronic, magnetic, and topological phenomena. Understanding the consequence of polarity mismatch between the substrate and the DPO would be the first step towards this broad goal. We investigate this by studying the interface between a prototypical insulating DPO Nd$_2$NiMnO$_6$ and a wide-band gap insulator SrTiO$_3$. The interface is found to be insulating in nature.
By combining several experimental techniques and density functional theory, we establish a site-selective charge compensation process that occurs explicitly at the Mn site of the film, leaving the Ni sites inert. We further demonstrate that such surprising selectivity, which cannot be explained by existing mechanisms of polarity compensation, is directly associated with their electronic correlation energy scales. This study establishes the crucial role of Mott physics in polar compensation process and paves the way for designer doping strategies in complex oxides. 
\end{abstract}

\maketitle

 Understanding the consequences of electrostatic boundary conditions at the interface between two dissimilar materials has been a long-standing and ever-explorative question in condensed matter physics~\cite{tsymbal:2012}. Semiconductor heterostructures became a starting ground to study the effect of polarity mismatch, also termed as polar catastrophe, wherein Ge/GaAs heterostructure came to be studied in great detail~\cite{Frensley:1977p2642}. Although both charge and atomic rearrangements were theoretically demonstrated as feasible ways to compensate the diverging potential~\cite{Frensley:1977p2642,Harrison:1978p4402}, no electronic reconstruction was observed experimentally~\cite{Frensly:1977p48,Grant:1978p656}, which can be attributed to the lack of multiple valence states of elements present in a typical semiconductor heterostructure~\cite{Pollmann:1982p257}. 
From this aspect, complex oxide heterostructures are fundamentally different as transition metal (TM) ions can easily access multiple valence states and have strong interplay amongst spin, charge, lattice, and orbital degrees of freedom~\cite{Hwang:2012p103,Chakhalian:2014p1189,Middey:2016p305}. In recent years, the impact of polarity mismatch in $AB$O$_3$ based perovskite oxide heterostructures has become a research theme in itself, wherein a plethora of emergent phenomena such as two-dimensional electron gas ~\cite{Ohtomo:2004p423,Stemmer:2014p151}, superconductivity~\cite{Reyren:2007p1196}, ferromagnetism~\cite{ Bert:2011p767,Li:2011p762,Lee:2013p703}, superparamagnetism~\cite{Anahory:2016p12566}, etc. have been demonstrated. Unlike the case in a semiconductor, the potential divergence can be avoided here through either electronic/chemical reconstruction, resulting in a change in cationic valency~\cite{Nakagawa:2006p204,Tazikawa:2009p236401,Chen:2017p156801,Park:2013p017401,Middey:2014p6819,Mundy:2014p3464,Yu:2014p5118} or structural effect such as interfacial intermixing~\cite{Chambers:2011p206802,Saluzzo:2013p2333}, etc. 

Double perovskite oxides ($A_2$\emph{BB’}O$_6$) where \emph{B}, \emph{B’} are the TM elements, exhibit diverse phenomena including high-temperature ferromagnetism~\cite{Sarma:2000p2549}, room temperature magnetoresistance~\cite{Kobayashi:1998p667}, multiglass behavior~\cite{Choudhury:2012p127201}, insulating ferromagnetism~\cite{Rogado:2005p2225,Saha:2020p014003}, etc. with lots of potential for technological applications~\cite{Vasala:2015p1}.  There are also predictions of realizing topological phases in DPOs through heterostructuring routes~\cite{Cook:2014p077203}. Surprisingly, the ways of polar compensation in the ultrathin film of DPOs have rarely been addressed~\cite{Spurgeon:2018p134110,DeLuca:2022p2203071}. This question is further connected with a more fundamental question about the role of Mott physics for polar compensation in  oxide heterostructures as TM oxides, specifically 3\emph{d} TMOs,  are strongly correlated electron systems. 
Our hypothesis is if strong correlation physics is crucial in contrast to the conventional semiconductor-like band bending framework, $B$ and $B'$ would respond differently to the polar catastrophe due to their different characteristic correlated energy scales like the on-site coulomb repulsion energy ($U$), charge transfer energy ($\Delta$) and the hopping parameter ($t$)~\cite{Zaanen:1985v55}. 

In this work,  we have investigated thin films of   Nd$_2$NiMnO$_6$ (NNMO), grown on single crystalline NdGaO$_3$ (NGO) (110)$_\mathrm{or}$ [or represents orthorhombic coordinate system] and SrTiO$_3$ (STO) (001) substrates by pulsed laser deposition (PLD) technique. Bulk NNMO is a monoclinic ferromagnetic insulator with a Curie temperature of 200 K~\cite{Pal:2019p045122}. Contrary to the NNMO film on NGO where both film and substrate consist of alternating +1 and -1 charged plane (hence no interfacial polarity mismatch), the NNMO/STO heterostructure encompasses interfacial polarity mismatch, similar to the well-known LaAlO$_3$/SrTiO$_3$ (LAO/STO) heterostructure~\cite{Ohtomo:2004p423}. We have explored the effect of polar catastrophe using element-sensitive X-ray absorption spectroscopy (XAS) and scanning transmission electron microscopy (STEM) + electron energy loss spectroscopy (EELS) measurements along with first principles calculations. Our findings establish that polar compensation occurs through the formation of Mn$^{3+}$ within the film side near the interface, while the Ni site remains unperturbed. The additional electron comes due to surface oxygen vacancies. This site-selective charge compensation emanates from the relative difference between $U$ and $\Delta$ of Ni and Mn, demonstrating a direct connection between polarity compensation and electronic structure parameters in complex oxide heterostructures.
 
\begin{figure}[] 
	\vspace{-2pt}
	\hspace{-2pt}
	\includegraphics[width=0.8\textwidth] {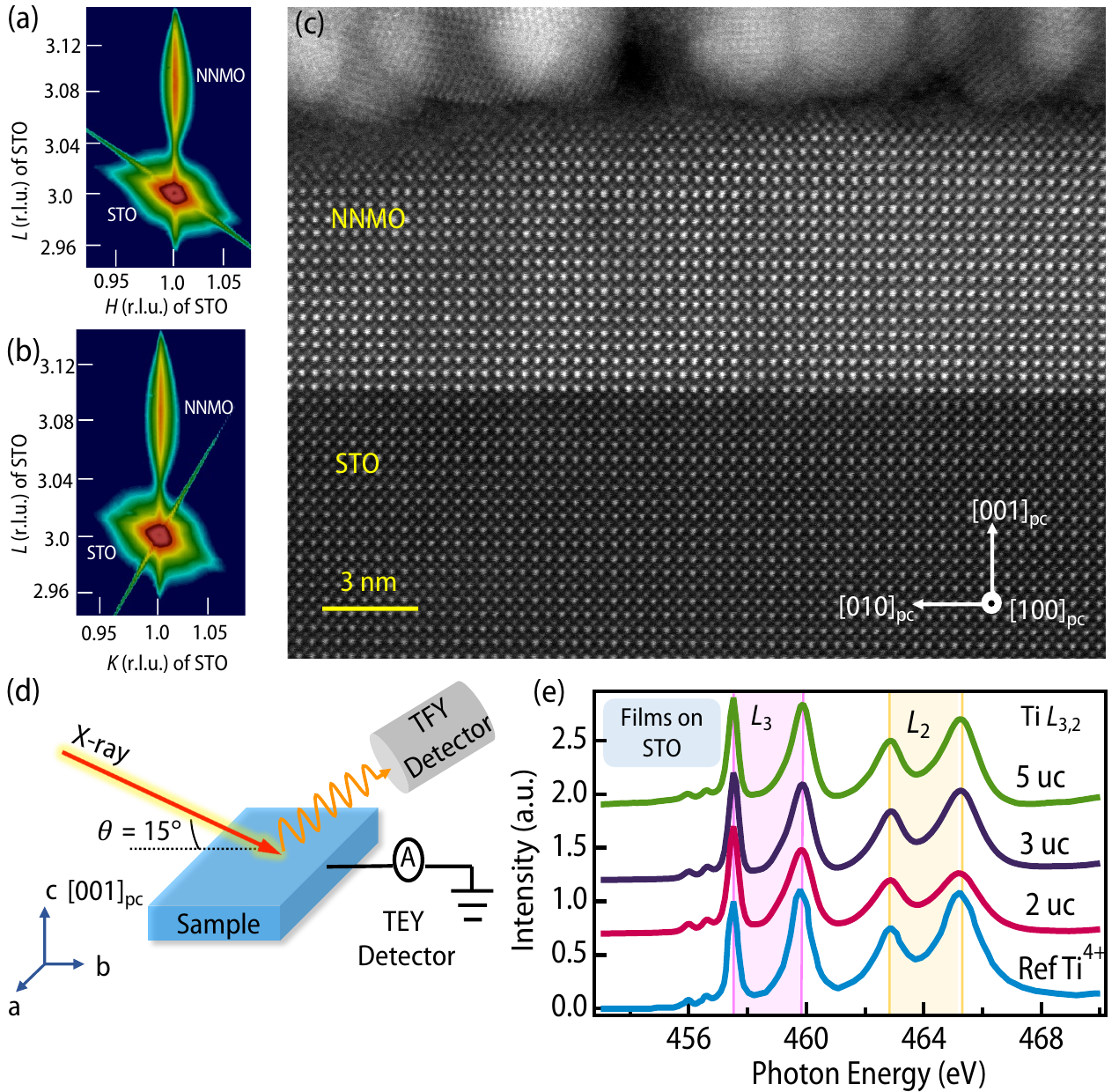}
	\caption{\label{Fig1} Reciprocal space mapping (RSM) in (a) $H-L$ plane (b) $K-L$ plane for a 20 uc film on STO around (113)$_\mathrm{pc}$ Bragg reflection. (c) HAADF-STEM image of the film.  (d) Schematic of the XAS measurement setup. XAS data recorded at 300 K at (e) Ti \emph{L}$_{3,2}$ edge for 2, 3 and 5 uc films on STO, compared with standard Ti$^{4+}$ spectra of STO.}
\end{figure}

NNMO films of several thicknesses [2, 3, 5, 10, and 20 uc, uc= unit cell in pseudocubic ($\mathrm{pc}$) notation] were grown on TiO$_{2}$ terminated STO (001) substrates in layer-by-layer fashion using a PLD system, connected with a high-pressure reflection high energy electron diffraction (RHEED) setup. 5, 10, and 20 uc NNMO films were also grown on GaO$_{2}$ terminated NGO (110)$_\mathrm{or}$ [(001)$_\mathrm{pc}$] substrates. The epitaxial strain is calculated to be +0.4\% and +1.5\% for NNMO on NGO, and STO, respectively. Fig.~\ref{Fig1}(a) and (b) show reciprocal space map (RSM) for the 20 uc NNMO film on STO in $H-L$ and  $K-L$ plane around the (113)$_\mathrm{pc}$ Bragg reflection of the substrate, recorded with synchrotron X-ray. The observation of same $H$ and $K$ value for the film and substrate confirms that the film is epitaxially strained with the substrate. Using this RSM image, the out-of-plane lattice constant ($c_\mathrm{pc}$) is found to be 3.803 \AA, very similar to the value found from a (00$L$) crystal truncation rod measurement. 

The epitaxial growth has been further confirmed by a cross-sectional STEM experiment on a 20 uc film on STO. High-angle annular dark field (HAADF) image along the zone axis [100] confirmed coherent epitaxial growth of the film, devoid of any defect or dislocation [Fig.~\ref{Fig1}(c)]. Atomically resolved electron energy loss spectroscopy (EELS) mapping shows that the extent of cationic intermixing across the interface is limited to 1 uc on the film side, also consistent with the film/substrate interface roughness obtained from XRR fitting. Since minuscule intermixing, likely to be inherent to all films due to same growth condition, cannot explain the systematic electronic structure changes with film thickness, shown in latter part of the paper. Another notable observation is that we do not find any NiO precipitation, which was reported to be the polar compensation mechanism for a thick ($\sim$ 40 nm) La$_2$NiMnO$_6$ film~\cite{Spurgeon:2018p134110}.
 
The smaller band gap of NNMO ($\sim$1-1.5 eV) ~\cite{Pal:2019p045122,nasir:2019p141} compared to STO ($\sim$3.25 eV) should result in a straddled band alignment, similar to the well-studied LaMnO$_3$/SrTiO$_3$ (LMO/STO) heterostructure~\cite{Chen:2017p156801,Anahory:2016p12566,Kasper:2019p1801428,Garcia:2010p627,Garcia-Barriocanal:2010p82,Wang:2015p716}. While for LMO/STO, the electronic reconstruction should occur within the LMO film ~\cite{Chen:2017p156801,Kasper:2019p1801428}, there are also reports of charge compensation within the STO substrates~\cite{Garcia:2010p627,Garcia-Barriocanal:2010p82}. Therefore, it is crucial to probe the electronic structures of both film and substrate for our heterostructures. For checking the valence state of the TM sites, we have performed element-sensitive XAS experiments for all samples grown on both substrates~\cite{Stohr:2006book}. All XAS data discussed in the main text were recorded in surface-sensitive total electron yield (TEY) mode with a grazing incidence geometry (Fig.~\ref{Fig1}(d)).

The origin of the metallic interface for LAO/STO heterostructure is often attributed to the polar catastrophe-driven electronic reconstruction, resulting in Ti$^{3+}$ in STO near the interface~\cite{Nakagawa:2006p204} or oxygen vacancy in STO leading to electron doping in Ti~\cite{Herranz:2007p216803}. Our measurements on NNMO/STO films confirm that the spectral features of Ti \emph{L}$_{3,2}$ edges are very similar to Ti$^{4+}$ of bulk stoichiometric STO [Fig.~\ref{Fig1}(e)].
Further peak analysis of the \emph{L}$_{2}$ and \emph{L}$_{3}$ edges, known to detect even small amount of Ti$^{3+}$ ions~\cite{Garcia-Barriocanal:2010p82} do not find any signature of electron doping on Ti. (also supported by Ti \emph{L}$_{3,2}$ edge EELS).These results are further corroborated by electrical transport measurements with wire-bonded contacts, which do not find any measurable interfacial conductivity,  similar to LaMnO$_3$/SrTiO$_3$~\cite{Chen:2017p156801}, LaCrO$_3$/SrTiO$_3$ heterostructures~\cite{Chambers:2011p206802}.The presence of any bulk oxygen vacancy within the substrates can also be discarded as all films were annealed in-situ for 30 minutes under 500 Torr ultrapure oxygen atmosphere post deposition. 

\begin{figure}[] 

	\vspace{-3pt}
	\hspace{-6pt}
	\includegraphics[width=0.8\textwidth] {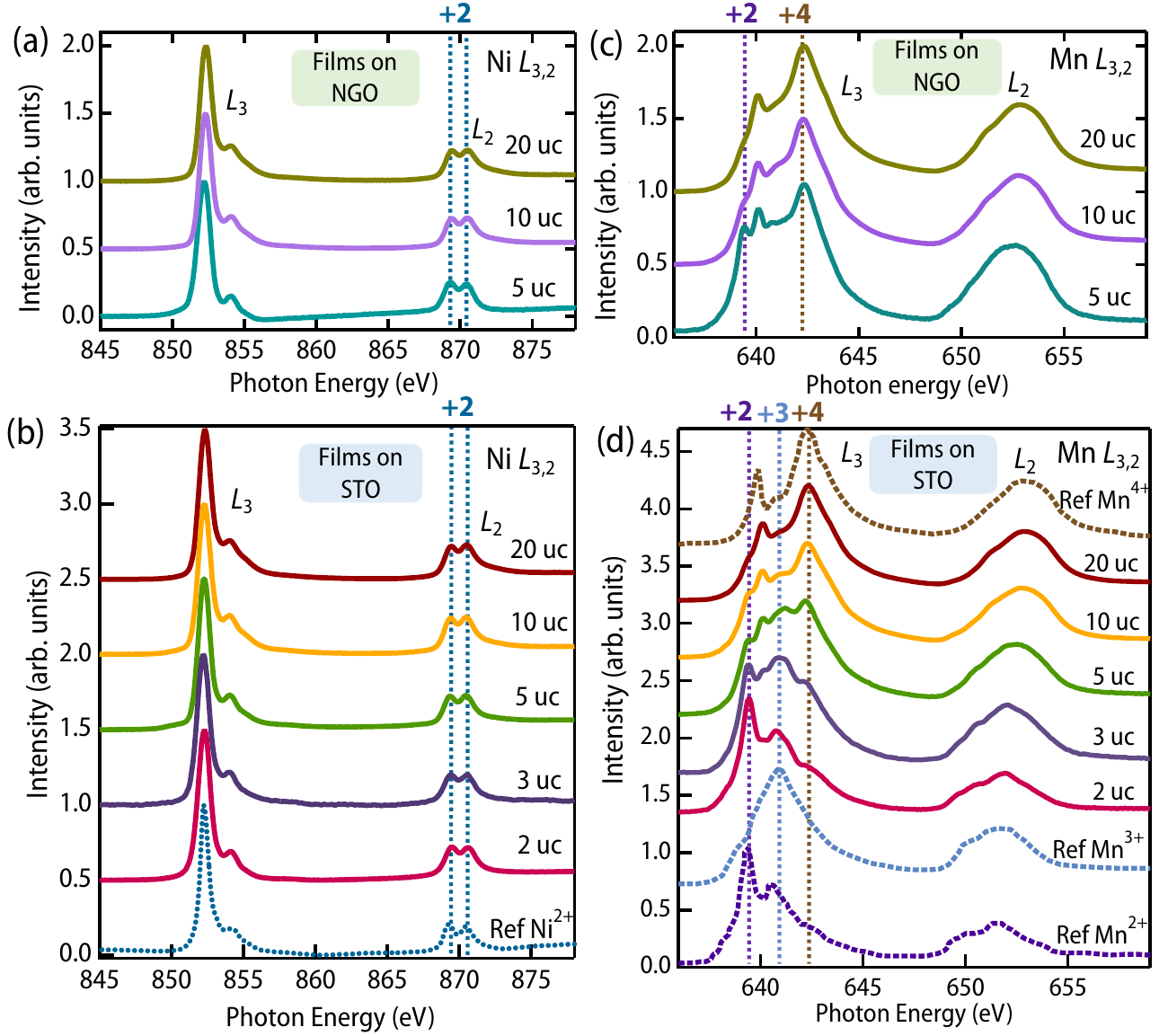}
	\caption{\label{Fig2} XAS data recorded at 300 K at (a) Ni \emph{L}$_{3,2}$ edge for 5, 10 and 20 uc films on NGO (b) Ni \emph{L}$_{3,2}$ edge for 2, 3, 5, 10 and 20 uc films on STO. Reference for Ni$^{2+}$ spectrum adapted from Ref.  [\onlinecite{Morrow:2016p3666}]. (c) Mn \emph{L}$_{3,2}$ edge for 5, 10 and 20 uc films on NGO (d) Mn \emph{L}$_{3,2}$ edge for 2, 3, 5, 10 and 20 uc films on STO. References Mn$^{2+}$, Mn$^{3+}$ and Mn$^{4+}$ adapted from Ref. [\onlinecite{Pal:2018p165137}]}. 
\end{figure}

In the absence of any charge compensation in STO substrate, we proceed to the film side. In Fig.~\ref{Fig2}(a),(b) we show Ni $L_{3,2}$ edge XAS spectra of several films on NGO and STO along with a Ni$^{2+}$ standard~\cite{Morrow:2016p3666}. Admittedly, Ni remains in +2 state for all films on both substrates, akin to bulk NNMO~\cite{Pal:2019p045122}.  EELS analysis further verifies this, asserting that the polar mismatch does not affect Ni ions.

 In sharp contrast, we find a systematic change in Mn XAS spectra [Fig.~\ref{Fig2}(d)]. For easy comparison, Mn XAS spectra for the standard of Mn$^{2+}$, Mn$^{3+}$ and Mn$^{4+}$  have also been shown in Fig.~\ref{Fig2}(d). The spectra for films on NGO (no polar mismatch) [Fig.~\ref{Fig2}(c)] confirm Mn$^{4+}$ state, similar to bulk NNMO~\cite{Pal:2019p045122}. However, a small feature of Mn$^{2+}$ around 639.4 eV is present, becoming much more prominent for the 5 uc film. On the contrary, there is an additional change in Mn XAS for the films on STO(Fig.~\ref{Fig2}(d)). Apart from the Mn$^{2+}$ feature also found here with a similar trend with film thickness, an additional feature for Mn$^{3+}$ states is observed, increasing on lowering the film thickness. These Mn$^{3+}$ signatures become comparable to Mn$^{4+}$ for the 5 uc film and even become the dominant contributor for the 3 uc film. Bulk-sensitive total fluorescence yield (TFY) data also verifies the same trend. Notably, surface oxygen vacancies leading to Mn$^{2+}$ ions are common for manganite thin films~\cite{Calderon:1999p6698,Kasper:2019p1801428,Li:2012p4312,Tebano:2008p137401,Pesquera:2012p1189} and has also been designated to provide electrons for polarity compensation in LMO/STO heterostructures~\cite{Kasper:2019p1801428}. However, such surface Mn$^{2+}$ cannot compensate for polar catastrophe here as it is present in films on both STO and NGO. Rather, the compensation is happening through the formation of Mn$^{3+}$, being exclusively present only for films on STO. We also note that TEY mode probes $\sim$ 5 nm length scale from the surface and contribution of the deeper layers decreases exponentially from the surface~\cite{Stohr:2006book}. Thus, the decrease of Mn$^{3+}$ feature with the increase in NNMO film thickness indicates non-uniformity of Mn valence along the out-of-plane film direction with Mn$^{3+}$ hosted in a few layers. 
  
  \begin{figure}[]
  
	\centering
	{~} \vspace{-2pt}
	{~} \hspace{0pt}
	\includegraphics[width=0.8\textwidth] {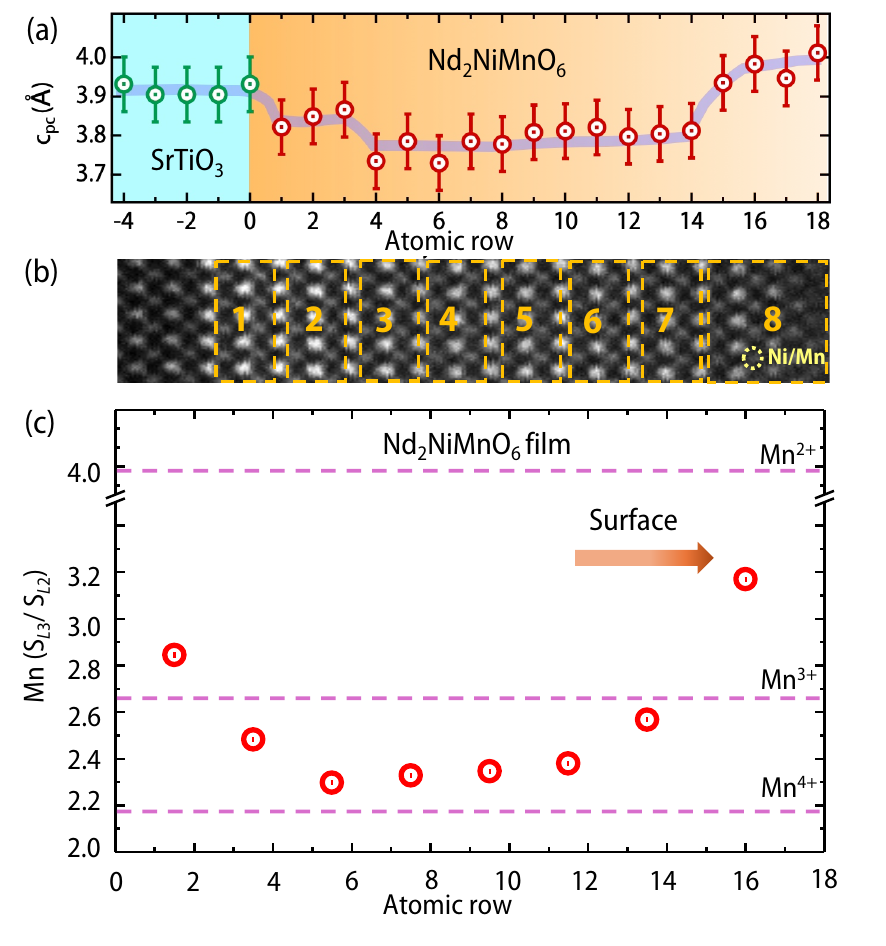}
	\caption{\label{Fig3}(a) Layer resolved $c_\mathrm{pc}$ for 20 uc NNMO on STO, evaluated from HAADF-STEM image. (b) EELS spectra have been averaged region-wise, marked by 1 to 8.(c)  Mn S$_{L3}$/S$_{L2}$ plotted region-wise. The dashed lines correspond to  Mn S$_{L3}$/S$_{L2}$ references for Mn$^{2+}$, Mn$^{3+}$ and Mn$^{4+}$.}
	
\end{figure}
The ionic radii of Mn$^{2+}$, Mn$^{3+}$ and Mn$^{4+}$  in octahedral environment with high spin configuration are 0.83 \AA, 0.645 \AA, and 0.53 \AA, respectively~\cite{Shannon:1976p751}.
Thus, the layer-resolved $c_\mathrm{pc}$ would vary
according to the Mn oxidation state if present in different layers.
Fig.~\ref{Fig3}(a) shows the variation of the layer resolved $c_\mathrm{pc}$, evaluated using HAADF-STEM image, demonstrating that the film has three types of layers: $c_\mathrm{pc}$ for intermediate layers (4-14 uc from the interface) are similar to the value obtained from RSM measurement (Fig.~\ref{Fig1}(a),(b)), few outermost layers  have higher value of $c_\mathrm{pc}$ validating  Mn$^{2+}$ near film surface and most importantly, the $c_\mathrm{pc}$ value for the first 3 uc of the film is slightly higher, compared to the intermediate layers, implying that 2-3 uc near the interface regions are enriched with Mn$^{3+}$. This aspect has been further tested by Mn EELS as the ratio of the area under $L_3$ and $L_2$ edge (S$_{L3}$/S$_{L2}$) strongly depends on the Mn oxidation state~\cite{Leapman:1980p397,Varela:2009p085117,Bawane:2021p113790}. Similar analysis of Mn EELS recorded from different regions of the 20 uc NNMO film on STO clearly demonstrate the interfacial regions are enriched with Mn$^{3+}$, whereas surface layers have Mn$^{2+}$ [Fig.~\ref{Fig3}(b)-(c)].
We also performed annular bright field imaging (ABF-STEM), sensitive to the oxygen atoms, revealing reduced oxygen octahedral distortion near the interface, in line with Mn$^{3+}$ enrichment, compared to the more distorted bulk Mn$^{4+}$ layers. Overall, our comprehensive experiments demonstrate that NNMO films on STO exhibit site-selective electron doping at interfacial Mn states, accompanied by modulations in octahedral rotations on both sides of the interface.

 \begin{figure}[] 
	\vspace{-1pt}
	\hspace{0pt}
	\includegraphics[width=0.8\textwidth] {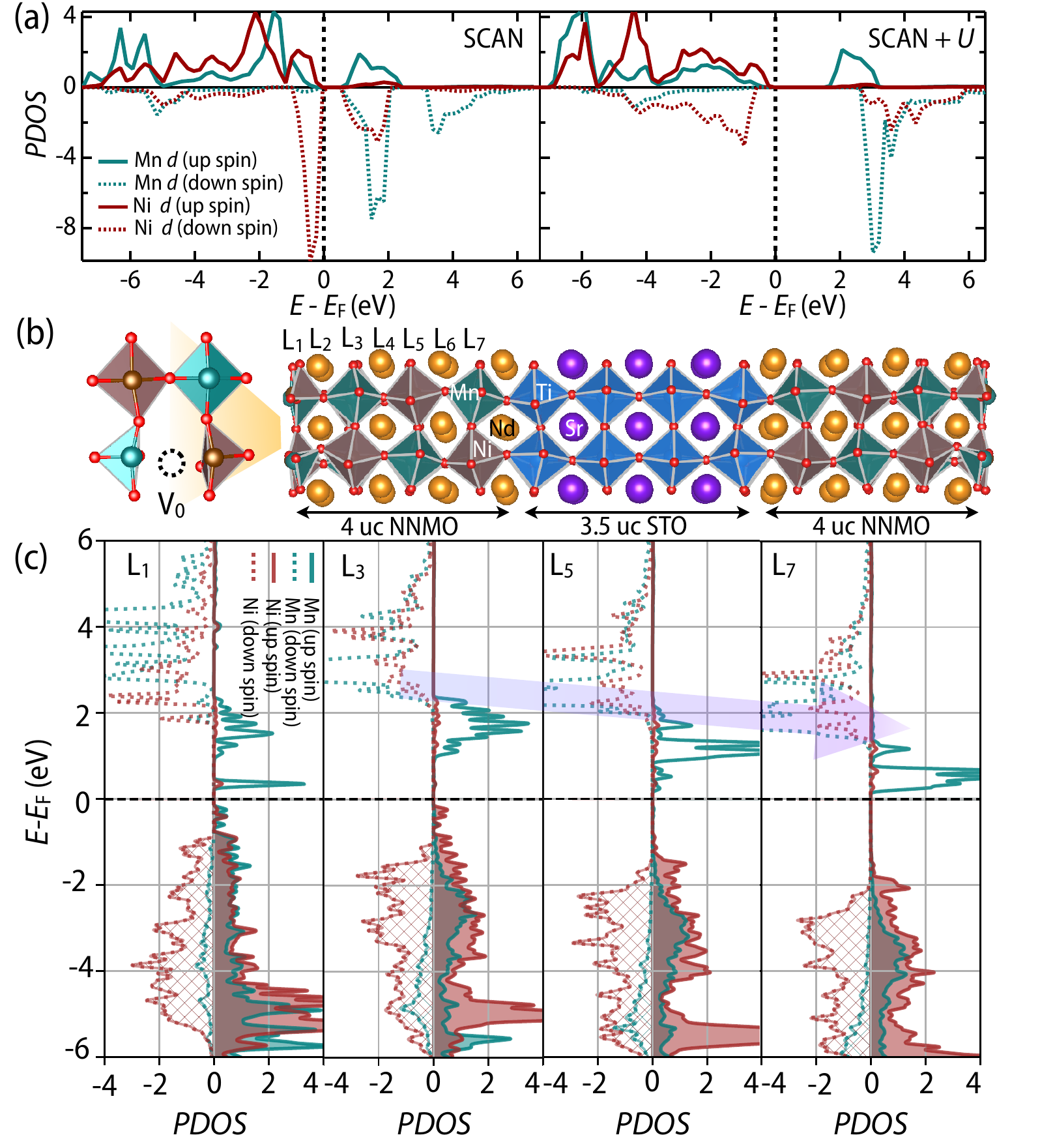}
	\caption{\label{Fig4}(a) Projected density of states ($PDOS$) for Ni and Mn $d$ states for bulk NNMO using SCAN (left), SCAN+$U$ (right). (b) Structure of 4 uc NNMO/ 3.5 uc STO/ 4 uc NNMO with one oxygen vacancy created on each surface of the NNMO slabs (marked as V$_0$), post relaxation used in DOS calculations. Top view of surface with the V$_0$ has also been shown. (c) Layer resolved projected density of states of Ni and Mn $d$ states using SCAN + $U$ functional corresponding to the layers marked in (b). Purple arrow showing shift of the Mn $d$ states towards Fermi is guide to the eye.}
\end{figure}

While, oxide interfaces with interfacial polar mismatch have been widely studied in the last 20 years, the novelty of our work lies in the selective participation of Mn and innertness of Ni towards compensation. To examine the role of on-site electron correlation ($U$), we performed first-principles DFT calculations for bulk NNMO using SCAN and SCAN+$U$. While in the SCAN framework the lowest unoccupied states ($E$ $\geq$ $E_F$) arise from both Ni and Mn $d$ states, SCAN+$U$ shows only Mn $d$ contribution [Fig.~\ref{Fig4}(a)]. To elucidate the polar catastrophe's impact on electronic structure, we performed DFT + $U$ calculations for a symmetric 4 uc NNMO/3.5 uc STO/4 uc NNMO slab using both PBE-sol and SCAN meta-GGA functionals \cite{PhysRevLett.115.036402, PhysRevB.54.11169, KRESSE199615}. To determine surface oxygen vacancy (V$_0$) effects, we simulated another slab with a surface V$_0$ [Fig.~\ref{Fig4}(b)]. The surface Mn magnetic moment is also enhanced to $\sim$ 3.96 $\mu_B$ which is considerably higher than the bulk Mn states ($\sim$ 3.2 $\mu_B$) corroborating the surface Mn$^{2+}$ ions. In both cases: pristine and surface V$_0$ [Fig.~\ref{Fig4}(c)], layer-resolved DOS confirm Mn 3$d$-dominated states are just above the Fermi level ($E_F$), while the Ni states remain higher in energy, precisely matching the experimentally observed site-selectivity. Moreover, the same analysis reveals that the minima of the conduction band move towards $E_F$ from L$_1$ (surface) to L$_7$ (interface) layers due to the internal electric field, estimated to 0.14 V/\AA\. Thus, a further reduction in interfacial conduction band energy in thicker slabs would lead to Mn$^{3+}$ at the interface.

  \begin{figure}[] 
	\vspace{-1pt}
	\hspace{0pt}
	\includegraphics[width=0.8\textwidth] {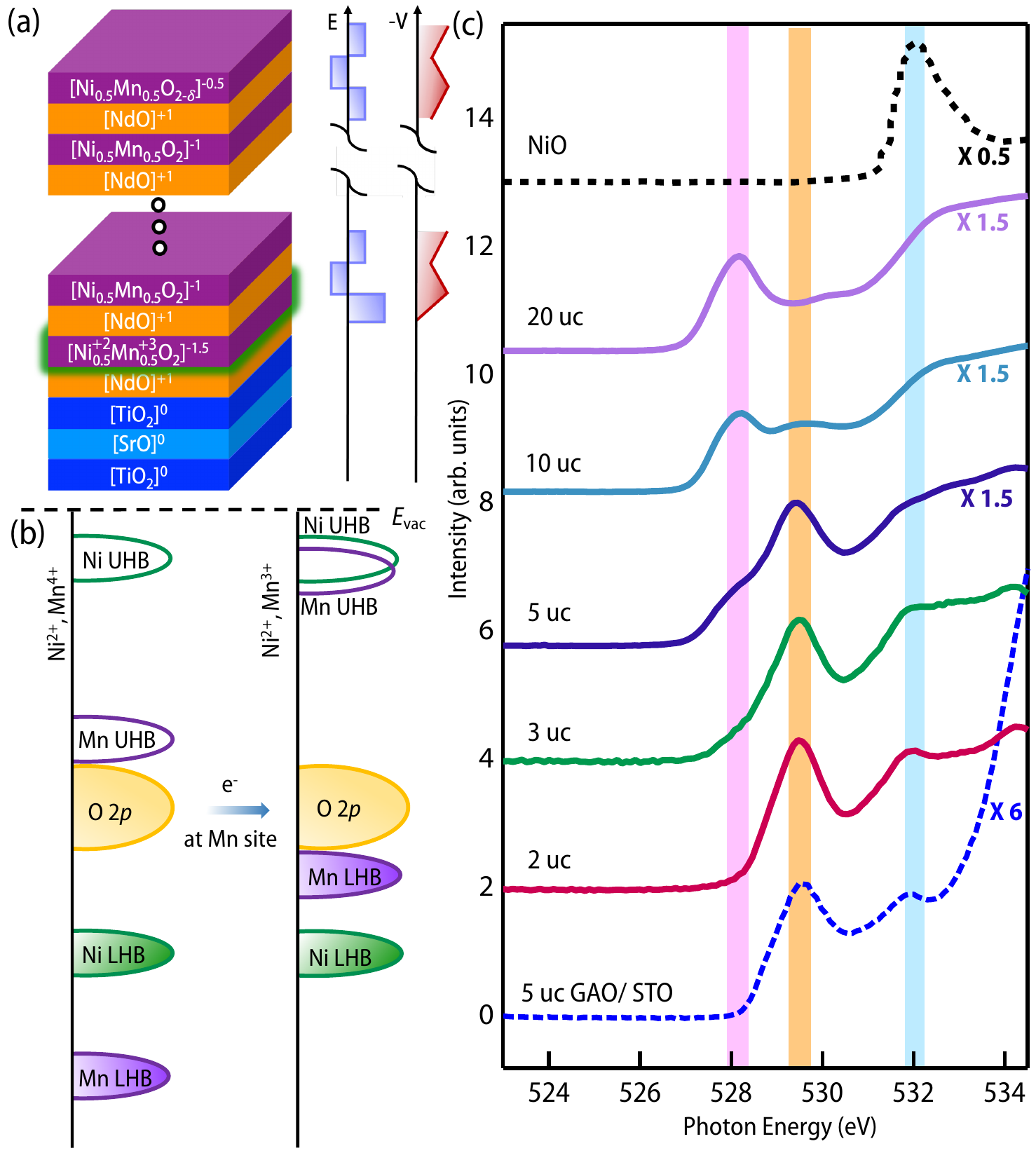}
	\caption{\label{Fig5}(a) Schematic of proposed charge compensation for NNMO on STO demonstrating transfer of 1/2 e$^{-}$ per uc to the interface (green highlighted region). (b) Energy band representation of UHB and LHB of Ni$^{2+}$ and Mn$^{4+}$ cations (left) and of Ni$^{2+}$ and Mn$^{3+}$ cations (right). (c) O \emph{K} edge XAS of 2, 3, 5, 10 and 20 uc films on STO at 300 K compared with reference of $\gamma$ Al$_{2}$O$_{3}$/STO[\onlinecite{Ojha:2021p054008}] and NiO[\onlinecite{Kuiper:1989p221}]. Pink shaded line represents Mn $e_g^\uparrow$-O 2$p$ and  Mn $t_{2g}^\downarrow$-O 2$p$, orange shaded line represents Ti $t_{2g}$ - O 2$p$, and blue shaded line represents Ni UHB - O 2$p$ and Ti $e_{g}$ - O 2$p$ contributions to the spectra, meant as guide to the eye.  }
\end{figure}

 Based on our results of XAS and STEM experiments, a charge compensation, schematically depicted in Fig.~\ref{Fig5}(a), can be considered to alleviate polar catastrophe via transfer of 1/2 electron per uc to the interfacial Mn forming Mn$^{3+}$ cations~\cite{Mundy:2014p3464} and surface oxygen vacancies act as electron donors. However, this conventional mechanism fails to explain why only the Mn site participates in this mechanism as the potential slope is independent of the nature of the cation. In order to explain such site selectivity, we invoke the physics of strong electron correlation for 3$d$ TMOs. In oxides, Ni$^{2+}$ has a typical $U\sim$ 6.5 eV and $\Delta \sim$ 4 eV [see page 102, Ref. ~\cite{Khomskii:2014book}]. The corresponding parameters for Mn$^{4+}$ are $\sim$ 5 eV and 1 eV, respectively. In Fig. ~\ref{Fig5}(b), we show the upper Hubbard band (UHB) and lower Hubbard band (LHB) for both Ni$^{2+}$ and Mn$^{4+}$ with a common O 2$p$ band. The Mn UHB is at an energetically lower value than Ni UHB, making it a preferred host to electron addition. The resultant energy level diagram after addition of an electron to the Mn is also depicted in Fig.~\ref{Fig5}(b), highlighting the significant overlap between UHBs of Ni$^{2+}$ and Mn$^{3+}$ [$U\sim$ 5 eV and $\Delta \sim$ 4 eV for Mn$^{3+}$, Ref. ~\cite{Khomskii:2014book}]. 

The above modification in unoccupied electronic states is verified by O $K$-edge XAS, which identifies relative cation energy positions above Fermi energy~\cite{Frati:2020p4056}.  
O $K$-edge XAS spectra for NNMO on STO are shown in Fig.~\ref{Fig5}(c) along with reference features related to Ti and Ni states ~\cite{Ojha:2021p054008, Kuiper:1989p221}. The feature of Ti $t_{2g}$ - O 2$p$ hybridized states of STO substrates are strongly present in the spectra of 5 uc film along with the films with lower thicknesses, and is weakly visible for the 10 uc film. The feature related to Ni UHB states appears around 532 eV for all NNMO films~\cite{Kuiper:1989p221}. The most notable observation is the evolution of feature around 528 eV with film thickness, which is contributed by the Mn $e_g^\uparrow$-O 2$p$ and  Mn $t_{2g}^\downarrow$-O 2$p$ states ~\cite{Galdi:2012p125129}. The gradual suppression of this peak with decreasing film thickness establishes the shift of Mn UHB to higher energy, as depicted in Fig.~\ref{Fig5}(b), due to charge compensation on Mn sites near the NNMO/STO interface. This is further supported by our observation that the 528 eV feature remains unaltered with the lowering of NNMO thickness grown on NGO.

To summarize, the present study unveils a surprising site-selective charge compensation in Nd$_2$NiMnO$_6$ thin films on SrTiO$_3$ (001) substrates via electron transfer from surface to the interfacial Mn states resulting in Mn$^{3+}$ cations at the interface. Despite being uniformly distributed as Mn, Ni sites remain unaffected by the process. By using O \emph{K} edge XAS, we further demonstrate that this site-selectivity is related to the relative energy position of the upper Hubbard band of Mn$^{4+}$ w.r.t. the Ni$^{2+}$. Thus, our results provide direct experimental evidence of strongly correlated electron physics' crucial role and inevitability in understanding polar compensation for oxide heterostructures. Further, exploring such selective doping will be an exciting avenue to realize exotic quantum phases in complex oxides.
\section{Acknowledgement}
The authors acknowledge XRD and wire bonding facilities of the Department of Physics, IISc Bangalore. SM  acknowledges funding support from a SERB Core Research grant (Grant No. CRG/2022/001906). SM also acknowledges the seed funding from Quantum Research Park (QuRP), funded by Karnataka Innovation and Technology Society (KITS), K-Tech, Government of Karnataka. SM also acknowledges funding from a DST Nano Mission consortium project [DST/NM/TUE/QM-5/2019]. NB acknowledges funding from the Prime Minister’s Research Fellowship (PMRF), MoE, Government of India. A. Sinha acknowledges JNCASR for providing a research fellowship and Param Yukti computing facility, NSM, JNCASR for providing computational resources. JM acknowledges UGC, India for fellowship. This research used resources of the Advanced Photon Source, a U.S. Department of Energy Office of Science User Facility operated by Argonne National Laboratory under Contract No. DE-AC02-06CH11357. This research used resources of the Advanced Light Source, which is a Department of Energy Office of Science User Facility under Contract No. DE-AC02-05CH11231. ZY acknowledges the National Natural Science Foundation of China (Grant No. 52202235).

\end{document}